\documentstyle[preprint,aps,epsf]{revtex} 

\begin{document} 

\title{Quantum Mechanics with Random Imaginary Scalar Potential}

\author{A. V. Izyumov and B. D. Simons}

\address{Cavendish Laboratory, Madingley Road, Cambridge, CB3 0HE, UK}

\maketitle

\begin{abstract}
We study spectral properties of a non-Hermitian Hamiltonian describing a 
quantum particle propagating in a random imaginary scalar potential. Cast in 
the form of an effective field theory, we obtain an analytical expression for 
the ensemble averaged one-particle Green function from which we obtain the 
density of complex eigenvalues. Based on the connection between non-Hermitian 
quantum mechanics and the statistical mechanics of polymer chains, we 
determine the distribution function of a self-interacting polymer in 
dimensions $d>4$. 
\end{abstract}

\newpage

The field of non-Hermitian quantum mechanics has, in recent years, attracted 
great interest. A variety of applications have been identified including the 
study of anomalous diffusion in random media~\cite{Isichenko92}, scattering
in open quantum systems~\cite{Sokolov88}, neural 
networks~\cite{Sompolinsky88}, chiral symmetry breaking in quantum 
chromodynamics~\cite{Stepanov96,Verbaarschot}, and the statistical mechanics 
of flux lines in superconductors~\cite{Hatano96}. The last of these has 
involved the study of the quantum mechanics of a particle confined to a 
random impurity potential and subject to an imaginary {\em vector} 
potential. This work revealed a novel mechanism of ``delocalisation'' of the 
quantum particle, sharply contrasting with the behaviour of the Hermitian 
counterpart~\cite{Hatano96,Feinberg97,Brouwer97,Janik97,Silvestrov98,Mudry98,%
Gade}. Here we investigate the 
spectral properties of a quantum particle confined to an imaginary {\em 
scalar} potential. To motivate our investigation, we apply these results to 
the study of the statistical mechanics of a self-interacting polymer chain.

The Hamiltonian describing a 
particle propagating in a random scalar potential is defined by
\begin{eqnarray}
\hat{H}={\hat{\bf p}^2\over 2m}+iV({\bf r}),
\label{hamil}
\end{eqnarray}
where the 
potential $V$ is drawn from a Gaussian distribution with zero mean, and 
correlator $\langle V({\bf r}) V({\bf r}^\prime)\rangle_V=\gamma
\delta^d({\bf r}-{\bf r}^\prime)$. To help motivate our discussion, we will
apply our analysis to the statistical mechanics of a polymer chain. 

In the continuum limit, the probability distribution $Z({\bf r},t)$ of the 
end-to-end distance ${\bf r}$ of a self-interacting polymer chain of length
$t$ can be expressed in the form of a path integral~\cite{Edwards,Kleinert},
\begin{eqnarray} 
Z({\bf r},t)=\int_{{\bf x}(0)={\bf 0}}^{{\bf x}(t)={\bf r}} D{\bf x}(\tau)
\exp\left\{-\int_0^t d\tau \frac{m}{2}\left(\frac{d{\bf x}}{d\tau}\right)^2
\right.-\left. \int_0^t\int_0^t d\tau' d\tau'' \frac{\gamma}{2}
\delta^d({\bf x}(\tau')-{\bf x}(\tau''))\right\}.
\label{distdef}
\end{eqnarray}
The first term, the Wiener measure, determines the entropic contribution,
while the second term represents the repulsive local contact interaction 
between the monomers that make up the chain. Decoupling the interaction by 
means of a Hubbard-Stratonovich field, the distribution function can be 
identified with the ensemble average of the Feynman propagator of the 
Hamiltonian above~(\ref{hamil}). Defining $\hat{U}(t)=\exp[-t\hat{H}]$,
\begin{eqnarray}
U({\bf r},t)\equiv\left\langle {\bf r}|\hat{U}(t)|{\bf 0}\right\rangle=
\int_{{\bf x}(0)={\bf 0}}^{{\bf x}(t)={\bf r}} D{\bf x}(\tau)\exp
\left\{-\int_0^t d\tau\left[\frac{m}{2}\left(\frac{d{\bf x}}{d\tau}\right)^2-
iV({\bf x}(\tau))\right] \right\},
\end{eqnarray}
the distribution is obtained from the ensemble average $Z({\bf r},t)=
\langle U({\bf r},t)\rangle_V$.

Applying a spectral decomposition of the complex Green function
\begin{eqnarray}
\hat{g}(z)\equiv {1\over z-\hat{H}}=\sum_i |R_i\rangle {1\over z-z_i} \langle
L_i|,
\end{eqnarray}
where $|R_i\rangle$ and $\langle L_i|$ denote the right and left-hand 
eigenfunctions of $\hat{H}$, and $z_i$ denote the complex eigenvalues, the 
distribution function~(\ref{distdef}) takes the form 
\begin{eqnarray}
Z({\bf r},t)={1\over \pi} \int d^2z \exp[-tz] {\partial\over \partial z^*} 
\left\langle g({\bf r},0;z)\right\rangle_V,
\label{dist}
\end{eqnarray}
where the integration runs over the entire complex plane.
Previous studies of the spectral properties of weakly non-Hermitian operators 
have largely (although not exclusively) focussed on properties of random matrix
ensembles~\cite{Ginibre65,Girko85,Sommers88,Fyodorov97a}. Such 
studies~\cite{Sommers88} have emphasised the pitfalls of a diagrammatic 
analysis based on a perturbative expansion of $\hat{g}$ in the random 
potential. The problems are revealed by representing the density of complex 
eigenvalues through the operator identity
\begin{eqnarray}
\rho(z)\equiv\sum_i\delta^2(z-z_i)={1\over \pi}{\partial\over \partial z^*}
{\rm tr}\ \hat{g}(z),
\end{eqnarray}
wherein the Green function is shown to be {\em non-analytic} everywhere in 
which the density of states (DoS) is non-vanishing. To circumvent these 
difficulties, a 
representation has been introduced~\cite{Sommers88,Feinberg97} in which the 
complex Green function is expressed through an auxiliary {\em Hermitian} 
operator,
\begin{eqnarray}
\hat{G}^{-1}(\epsilon)\equiv \pmatrix{\epsilon & z-\hat{H}\cr z^*-
\hat{H}^\dagger &\epsilon\cr}=\epsilon+\left(x-{\hat{p}^2\over 2m}\right)
\sigma_1-(y-V)\sigma_2,
\label{gorkov}
\end{eqnarray}
where $z=x+iy$ and $\mbox{\boldmath$\sigma$}$ represent Pauli matrices. 
Making use of this construction, a relationship between $\hat{g}$ and the 
matrix Green function is straightforwardly obtained,
\begin{eqnarray}
\hat{g}(z)=\lim_{\epsilon\to -i0} \hat{G}_{21}(\epsilon).
\end{eqnarray}
At the same time, this representation manifests an implicit {\em time-reversal}
and {\em chiral} symmetry of the matrix Hamiltonian: $\sigma_3\hat{G}^{-1}(0)
\sigma_3=-\hat{G}^{-1}(0)$. Recently, this representation was successfully 
combined with diagrammatic perturbation theory to study the spectrum of 
the Fokker-Planck operator describing particles diffusing in a quenched random 
velocity field~\cite{Chalker97}. 

Cast in this form, we are able to apply standard field theoretic methods to 
obtain statistical properties of the single-particle Green function. Our 
approach is closely related to that recently developed by 
Efetov~\cite{Efetov97} (see also Ref.~\cite{Andreev94}) to investigate 
spectral properties of the random Schr\"odinger operator subject to an 
imaginary vector potential. This approach involves a generalisation of the 
supersymmetry method originally tailored to the description of disordered 
conductors~\cite{Efetovbook}. Being somewhat more technical than the 
diagrammatic perturbation theory employed in~\cite{Chalker97}, the use of 
the supersymmetry technique is nevertheless justified for the problem at hand. 
Indeed, although the diagrammatic approach reproduces the mean-field result, 
it does not take properly into account the existence of the massless Goldstone
modes (see below), and therefore can not be safely used if one wants to go 
beyond the saddle-point approximation.

The analysis begins by expressing the matrix Green function as a functional 
integral over $8$-component supervector fields $G_{\alpha\beta}({\bf 0},
{\bf r})=i\langle{\rm tr}[R^{-1} \Psi({\bf 0})\otimes \bar{\Psi}({\bf r})
\sigma_3 R\Sigma^{\beta\alpha}]\rangle_\Psi/4$, where
\begin{eqnarray}
\left\langle\cdots\right\rangle_\Psi=\int D[\bar{\Psi},\Psi]\left(\cdots
\right)\exp\left\{-{i\over 2}\int \bar{\Psi}\left[i\left(x-{\hat{p}^2\over 2m}
\right)\sigma_2+\left(V-y\right)-i0\sigma_3\right]\Psi\right\},
\end{eqnarray}
$R=\exp[-i\pi\sigma_1/4]$, and the infinitesimal imaginary part ensures 
convergence. Here we adopt a standard 
notation~\cite{Efetovbook} in which the fields $\Psi$, $\bar{\Psi}$ subdivide 
into a time-reversal ({\sc tr}), a Fermion-Boson ({\sc fb}), and a 
``spinor'' or matrix sector. $\Sigma^{\beta\alpha}$ is a 
$2\times 2$ matrix which projects on to the $\alpha\beta$ components in the 
spinor space. 

Expressed in this form, an ensemble average over the random impurity potential 
generates a quartic interaction of the fields which can be decoupled by the 
introduction of $8\times 8$ component supermatrix fields $Q$. Taking our 
notation from disordered conductors and defining 
$\tau^{-1}(x)=2\pi\gamma\nu(x)$, where $\nu(x)$ is the unperturbed DoS, we 
obtain
\begin{eqnarray}
\left\langle \exp\left[-{i\over 2}\int \bar{\Psi}V\Psi\right]\right\rangle_V
=\int DQ \exp\left[{1\over 4\tau}\int\left({\pi\nu\over 2}{\rm str} Q^2
-\bar{\Psi}Q\Psi\right)\right],
\end{eqnarray}
where ${\rm str}M=M_{\rm\sc ff}-M_{\rm\sc bb}$ represents the trace operation 
for supermatrices. The supermatrix fields $Q$ have an algebraic structure 
which reflects that of the dyadic product $\Psi\otimes\bar{\Psi}$. 
Integrating over the superfields $\Psi$, we obtain $\langle 
G_{\alpha\beta}({\bf 0},{\bf r})\rangle_V=-\langle {\rm tr}[R^{-1} 
{\cal G}({\bf 0},{\bf r})\sigma_3 R\Sigma^{\beta\alpha}]\rangle_Q
/4$ with
\begin{eqnarray}
\left\langle\cdots\right\rangle_Q=\int DQ(\cdots)\exp\left[\int {\rm str}
\left({\pi\nu\over 8\tau}Q^2-{1\over 2}\ln {\hat{\cal G}}^{-1}\right)\right],
\label{action}
\end{eqnarray}
where the supermatrix Green function is defined by
\begin{eqnarray}
\hat{\cal G}^{-1}=i\left(x-{\hat{p}^2\over 2m}\right)\sigma_2-y-{i\over 2\tau}
Q.
\end{eqnarray}

Further progress is possible only within a saddle-point approximation, 
which is controlled in the limit of weak disorder $[y,1/\tau]\ll x$. Minimising
the action~(\ref{action}) with respect to variations in $Q$, we obtain the
saddle-point equation
\begin{eqnarray}
Q({\bf r})=-{i\over \pi\nu}{\cal G}({\bf r},{\bf r}).
\label{spe}
\end{eqnarray}
The saddle-point solution is 
found from the ansatz that $Q$ is homogeneous in space, and diagonal in the 
{\sc tr} and {\sc fb} sector. Applying the parametrisation $Q=q_0+{\bf q}
\cdot\mbox{\boldmath $\sigma$}$, we obtain two solutions~\cite{foot1},
\begin{eqnarray}
q_0=\cases{iy\tau\cr i{\rm sgn}(y)\cr},\qquad q_1=q_2=0,\qquad q_3=
\cases{(1-y^2\tau^2)^{1/2} \cr 0. \cr}
\end{eqnarray}
Moreover, invariance of the Eq.~(\ref{spe}) under rotations $Q\to TQT^{-1}$ 
where $[T,\sigma_2]=0$, shows that the first solution of the 
saddle-point equation spans a degenerate manifold, Class CI in the classification of 
Ref.~\cite{Zirnbauer96}. Below, we will find that this 
solution corresponds to the non-analytic part of the Green function with a 
non-zero DoS. The second (non-degenerate) solution yields 
the analytic part of the Green function, and does not contribute to the DoS. 
The region in the complex plane where the first solution is stable defines 
the support of the spectrum. Straightforward stability analysis shows that 
the boundary of the spectrum is determined by the equation $y=1/\tau(x)$.

Expanding the action in slow fluctuations $T({\bf r})$ around the 
saddle-point solution we obtain the low energy effective action
\begin{eqnarray}
S[Q]=-{\pi\nu\over 8}\int d{\bf r} D(y)\ {\rm str}\left(\partial Q\right)^2,
\end{eqnarray}
where $Q=T\sigma_3 T^{-1}$, and $D(y)=(1-y^2\tau^2)D_0$, with $D_0=2x\tau (x)/md$, 
denotes the $y$-dependent classical diffusion constant. Thus, in contrast to 
a {\em real} random impurity potential, modes of density relaxation of the
matrix Hamiltonian are controlled by massless Goldstone modes of a 
supersymmetric non-linear $\sigma$-model of symmetry class CI with a 
diffusion constant which depends explicitly on $y$. 

Expanding the supermatrix Green function,
\begin{eqnarray}
{\cal G}({\bf 0},{\bf r})=-i\pi\nu(x)f_d({\bf r}) \left[q_0+q_3 Q({\bf r})
\right],
\label{gres}
\end{eqnarray}
where $f_d({\bf r})={\rm Im}G_0^-({\bf 0},{\bf r})/{\rm Im}G_0^-({\bf 0},
{\bf 0})$ and $G_0^-=(x-{\bf p}^2/2m-i/2\tau)^{-1}$, we obtain
\begin{eqnarray}
\left\langle G_{\alpha\beta}({\bf 0},{\bf r})\right\rangle=
i{\pi\nu\over 4}f_d({\bf r}) \int DQ\ {\rm str}\left[R^{-1} \left(q_0+
Q({\bf r})\right)\sigma_3 R\Sigma^{\beta\alpha}\right]\ e^{-S[Q]}.
\end{eqnarray}
Applied to the DoS, the projection matrix takes the form $\Sigma=(\sigma_1+
i\sigma_2)/2$. In this case the supermatrix degrees of freedom of $Q$ are
decoupled from the source, and the DoS is specified simply by the mean-field 
result
\begin{eqnarray}
\rho(z)=\cases{(4\pi\gamma)^{-1}, &$|y|<1/\tau(x)$,\cr 0 &$|y|>1/\tau(x)$,}
\end{eqnarray}
satisfying the sum rule $\int dy \rho(z)=\nu(x)$.
This result compares with that obtained for the corresponding random matrix 
ensemble~\cite{Fyodorov97a}, and contrasts that obtained for an imaginary 
vector potential~\cite{Efetov97}. In particular, since the DoS 
source does not couple to the effective action, the mean field estimate is 
unchanged by integration over $Q$. As a result, the complex eigenvalue 
density remains non-singular at $y=0$. 

We have computed numerically the eigenvalues of the two-dimensional lattice 
version of the Hamiltonian~(\ref{hamil}). Theory and simulation are compared in
fig.\ \ref{fig1}, for $50$ realisations of a $32 \times 32$ lattice with 
$\gamma=1$. The solid line represents the 
boundary of the spectrum as calculated in the saddle-point approximation, 
$y=\pm 1/\tau(x)=2\pi\nu(x)\gamma$, $-4<x<4$, where $\nu(x)$ is the exact 
DoS of the clean two-dimensional tight-binding model. The 
results show good agreement between theory and numerics, with deviations 
becoming significant only near the edges of the band, and in the vicinity of 
the band center. (Note that, in contrast to the Hermitian disordered 
Hamiltonian, the imaginary scalar potential {\em raises} the energy of the 
low lying states, an effect easily understood within the framework of second 
order perturbation theory. However, the localisation properties of such 
``Lifshitz tail'' states, as well as their sensitivity to optimal 
fluctuations of the impurity potential is a subtle question which lies 
beyond the scope of the present investigation. Secondly, a sublattice 
symmetry specific to the tight-binding Hamiltonian induces a reflection 
symmetry of the spectrum around $x=0$. This additional symmetry, which 
parallels that found with the imaginary vector potential, leads to the 
accumulation of a finite fraction of states along $x=0$, a phenomenon which 
has no counterpart in the continuum model.) We believe that the small 
deviations of numerics and theory away from the band edge and band centre 
can be ascribed to the influence of massive fluctuations of the matrix 
fields, i.e. those which violate the symmetry $[T({\bf r}),\sigma_2]=0$.

Turning to the self-interacting polymer chain discussed above, Eq.~(\ref{gres})
can be used to obtain the probability distribution $Z({\bf r},t)$. However, 
to do so, we should recall that the saddle-point approximation is justified 
in the limit $x\tau\gg 1$ which, setting $m=1$, translates to the following 
condition on the contact interaction: $\gamma x^{(d-4)/2} \ll 1$. The latter 
is satisfied as $x\rightarrow 0$ in dimensions higher than four, and 
$x\rightarrow\infty $ in dimensions lower than four. Equivalently, applied to 
the polymer chain, the analysis above applies at time scales 
$t\gg\gamma^{2/(d-4)}$ for $d>4$, and $t\ll (1/\gamma)^{2/(4-d)}$ for 
$d<4$ defining the upper critical dimension as $d_c=4$~\cite{Kleinert}.
Evaluation of Eq.~(\ref{dist}) for $d>4$ gives the power spectrum
\begin{eqnarray}
Z({\bf p}, t) = \frac{\sin (\omega({\bf p})t)}{\omega({\bf p})t}\, 
\exp [-p^2 t],\qquad \omega({\bf p})=\pi\gamma |{\bf p}|^{d-2}.
\end{eqnarray}

This result parallels that obtained by Chalker and Wang~\cite{Chalker97} for 
the time evolution of the particle density of the random Fokker-Planck 
operator. In particular, in the limit $t \rightarrow \infty$ it recovers 
diffusive behaviour $\langle r^2 \rangle \sim t$, while at short times, when 
the saddle-point approximation is uncontrolled, we have $\langle r^2 \rangle 
\sim (\gamma t)^{2/(d-2)}$. The long-time behaviour for $d<4$ is strongly 
affected by low-energy states which, being sensitive to optimum fluctuations
of the random impurity potential, is not accessible within the present 
framework.

In conclusion, applying conventional field theoretic methods, we have shown
that spectral properties of the non-Hermitian Hamiltonian describing a
particle propagating in an imaginary scalar potential are governed by a 
supersymmetric non-linear $\sigma$-model (of symmetry class 
CI~\cite{Zirnbauer96}). The latter has been used to obtain the complex 
eigenvalue density. As an application of these results, we have obtained the 
power spectrum of the probability distribution function of a self-interacting 
polymer chain. Although the analysis developed here has focussed on 
one-particle properties, an extension of the present approach to treat 
fluctuation phenomena is, at least formally, straightforward (see, for 
example, Ref.~\cite{Altland98}). 

{\em Acknowledgements}: We are grateful to Alexander Altland, John Chalker, 
Igor Lerner, Charles Offer, and Damian Taras-Semchuk
for illuminating discussions. One of us (A.V.I.) would like to acknowledge 
the financial support of Trinity College, Cambridge.

\begin{figure}

\caption{Complex eigenvalues taken from $50$ realisations of a $32\times 32$
square lattice tight-binding model with $\gamma=1.0$. The solid line 
is $y=\pm 1/\tau(x)=2\pi\nu(x)\gamma$.}
\label{fig1}

\end{figure}

\vskip1truein

\begin{figure}
\epsfxsize=15cm \epsffile{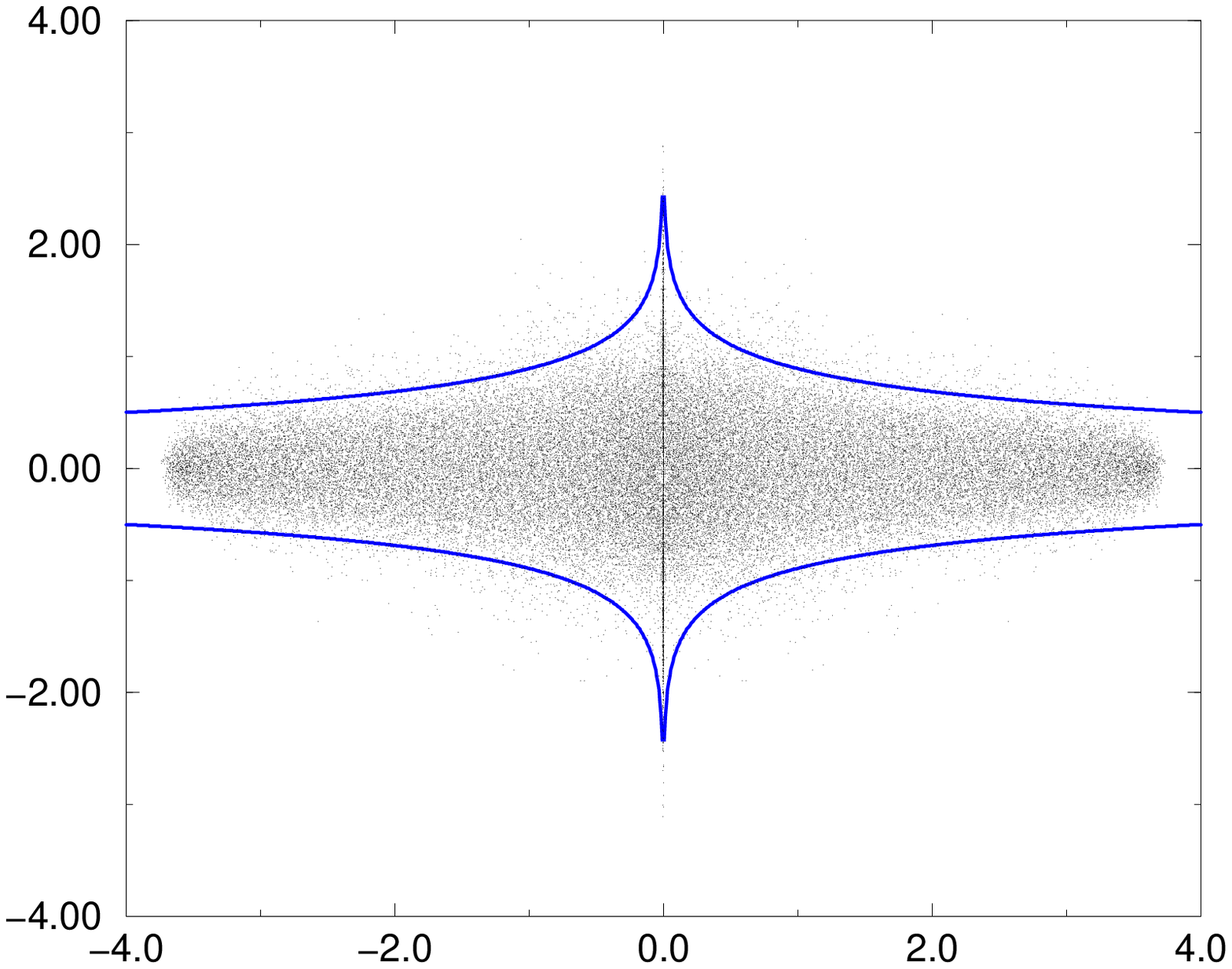}
\end{figure}


\begin{thebibliography}{99}
\bibitem{foot1}
According to the symmetry properties of $Q$~\cite{Efetovbook}, both 
saddle-point solutions are, strictly speaking, inaccessible. However, as 
shown in Ref.~\cite{Zirnbauer96} (see also \cite{Altland98}), the 
absence of singularities in the action allow for an analytic continuation of 
the integration contour outside the region admitted by the symmetry properties 
of $Q$.
\end{thebibliography}
\end{document}